\newcommand{\be}{\begin{equation}}
\newcommand{\ee}{\end{equation}}
\newcommand{\bea}{\begin{eqnarray}}
\newcommand{\eea}{\end{eqnarray}}
\newcommand{\beas}{\begin{eqnarray*}}
\newcommand{\eeas}{\end{eqnarray*}}

\newcommand{\slsh}[1]{{\not \! #1}}

\documentclass[twocolumn,aps,showpacs,nofootinbib,epsfig]{revtex4}
\usepackage{graphics}
\usepackage{epsfig}
\begin{document}
\title{Dynamical mass generation in strongly coupled Quantum Electrodynamics
       with  weak magnetic fields}
\author{Alejandro Ayala$^\dagger$, Adnan Bashir$^{\ddagger\,\dagger}$, Alfredo
Raya$^\dagger$, and Eduardo Rojas$^\dagger$}    
\affiliation{$^\dagger$Instituto de Ciencias Nucleares, Universidad
Nacional Aut\'onoma de M\'exico, Apartado Postal 70-543, M\'exico
Distrito Federal 04510, M\'exico.\\
$^\ddagger$Instituto de F{\'\i}sica y Matem\'aticas,
Universidad Michoacana de San Nicol\'as de Hidalgo, Apartado Postal
2-82, Morelia, Michoac\'an 58040, M\'exico.}

\begin{abstract}

We study the dynamical generation of masses for fundamental fermions in
quenched quantum electrodynamics in the presence of weak magnetic fields using
Schwinger-Dyson equations. Contrary to the case where the magnetic field is
strong, in the weak field limit the coupling should exceed certain critical
value in order for the generation of masses to take place, just as in the case
where no magnetic field is present. The weak field limit is defined as  $eB
\ll m(0)^2$, where $m(0)$ is the value of the dynamically generated mass in
the absence of the field.   We carry out a numerical analysis to study the
magnetic field dependence of the mass function above critical coupling and
show that in this regime the dynamically generated mass and the chiral 
condensate for the lowest Landau level increase proportionally to $(eB)^2$.
\end{abstract}


\maketitle

It is well known that in QED, fermions can acquire masses through self
interactions without the need of a nonzero bare mass. This phenomenon, known
as dynamical mass generation (DMG), happens above a certain critical value of
the coupling and its description can only be carried out in terms of
non-perturbative treatments. Schwinger-Dyson Equations (SDEs) provide a
natural platform to study DMG.  In the quenched version of QED, a favorite
starting point is to make an ansatz for the fermion-photon vertex and then
study the fermion propagator equation in its decoupled form.  It is also well
known that in the presence of strong magnetic fields, it is possible to
generate fermion masses for any value of the coupling. This phenomenon has
been given the name of {\it magnetic
  catalysis}~\cite{Gusynin,Leung,Hong,Ferrer}. Non-perturbative aspects of
dynamical mass generation in the presence of weak magnetic fields have earlier
been considered in the context of Nambu--Jona-Lasinio (NJL)
model~\cite{GusNJL}, QCD~\cite{Shushpanov} and (2+1)-dimensional
QED~\cite{Farakos}. In the context of QED4, the only work to our knowledge, is 
that of Kikuchi and Ng~\cite{Kikuchi}. However, they concentrate mainly on the
behaviour of the critical coupling in the presence of 
the weak magnetic fields. In this paper, we undertake the study
of the weak field dependence of the dynamically generated mass and the chiral
condensate in QED in the rainbow truncation of SDE.

The SDE for the fermion propagator without external fields (\emph{in vacuum})
in the rainbow approximation is 
\bea
   S_F^{-1}(p)=S_F^{(0)\,-1}(p)
- \frac{ i \alpha}{4\pi^3}  \int d^4k  \ \gamma^\mu S_F(k) 
   \gamma^\nu\Delta_{\mu\nu}^{(0)}(q),\label{SDE}
\eea
where $q\!=\!k\!-\!p$ and $\alpha=e^2/(4\pi)$ is the electromagnetic coupling
constant. In this expression  $\Delta_{\mu\nu}^{(0)}(q)$ is the bare photon
propagator, which in covariant gauges is written as
$\Delta_{\mu\nu}^{(0)}(q)=-\left(g_{\mu\nu}+(\xi-1)q_\mu q_\nu/q^2
\right)/q^2$, $\xi$ being the usual covariant gauge parameter. We write the
full fermion propagator as  
$S_F(p)={F(p^2)}/{(\slsh{p}-M(p^2))}$. $F(p^2)$ is referred to as the wave
function renormalization and $M(p^2)$ as the mass function. In the Landau
gauge $(\xi=0)$ $F(p^2)=1$ and the mass function has nontrivial solutions for
values of the coupling above the critical value $\alpha_c=\pi/3$. In the
presence of external fields, SDEs have been a subject of study already for
some time, see for example Ref.~\cite{gitman}.  

When the magnetic field is strong, Landau levels are separated from each other
by an amount $\sim \sqrt{eB}$ in such a way that for any  value of the
coupling $\alpha$, only the lowest Landau level (LLL) contributes to the
DMG~\cite{Gusynin,Leung,Hong,Ferrer}. However, in the case of  weak external
magnetic fields, Landau levels are close to each other and hence all 
contributions should be taken into account, which adds considerably to the
complexity of the problem, as emphasized also in 
the fourth article of reference~\cite{Gusynin}.

The presence of the field breaks Lorentz invariance. Consequently, a simple
Fourier transform on a single momentum variable is not possible. Nevertheless,
it has been shown~\cite{Ritus} that the mass operator in the presence of an
electromagnetic field can be written as a 
combination of the structures
\be
   \gamma^\mu\Pi_\mu\,,\, \sigma^{\mu\nu}F_{\mu\nu}\,,\,
   (F_{\mu\nu}\Pi^\nu)^2\,,\,\gamma_5F_{\mu\nu}\tilde{F}^{\mu\nu}
   \label{structures}
\ee
which commute with the operator $(\gamma\cdot\Pi)^2$, where 
$
   \Pi_\mu \!=\! i\partial_\mu \!-\! eA^{\mbox{\tiny{ext}}}_\mu,\
   F_{\mu\nu}\!=\!\partial_\mu A^{\mbox{\tiny{ext}}}_\nu \!-\!
              \partial_\nu A^{\mbox{\tiny{ext}}}_\mu,\
   \tilde{F}^{\mu\nu}=\frac{1}{2}\epsilon^{\mu\nu\lambda\tau}
   F_{\lambda\tau},\
   \sigma_{\mu\nu}=i[\gamma_\mu , \gamma_\nu]/2
$
and $A^{\mbox{\tiny{ext}}}$ is the external vector potential. We take 
$A^{\mbox{\tiny{ext}}}_\mu=B(0,-y/2,x/2,0)$ which describes a constant
magnetic field ${\mathbf{B}}=B{\mathbf{\hat{z}}}$~\cite{vanish}.  

In order to find a diagonal representation for the mass operator, we thus
need to find the eigenfunctions $\psi_{p\sigma}$ of the operator
$(\gamma\cdot\Pi)^2$, namely
\be
   (\gamma\cdot\Pi)^2\psi_{p\sigma}u_{\sigma\chi}=
   p^2\psi_{p\sigma}u_{\sigma\chi}
   \label{eigenp}\, ,
\ee
where $u_{\sigma\chi}$ are taken as the eigenspinors of $\Sigma_3$ and
$\gamma_5$. We work in cylindrical coordinates ${\bf r}=(r,\phi, z)$
and in the chiral representation of the $\gamma-$matrices where
$\Sigma_3$ and $\gamma_5$ are both diagonal with eigenvalues
$\sigma=\pm 1$ and $\chi=\pm 1$. The normalized eigenfunctions
$\psi_{p\sigma}$ (see for example Ref.~\cite{Sokolov}) are given by
\bea
   \psi_{p\sigma}(t,{\bf r})=N
   e^{-i(E_pt-p_zz)}
   e^{i(l_p-\frac{(\sigma + 1)}{2})\phi}
   I_{n_p-\frac{(\sigma +
   1)}{2}}^{s_p}(\rho)
\eea
where $N=\sqrt{2\gamma/(2\pi)^3}$, $\rho=\gamma r^2$,
$\gamma=eB/2$, $p^2=E_p^2-p_z^2-2eBn$ and  
\bea
   I_n^s(\rho)=\sqrt{\frac{s!}{n!}}e^{-\rho/2}\rho^{(n-s)/2}L_s^{n-s}(\rho)
   \label{Iexpl}
\eea
are the Laguerre functions~\cite{footnote1} with the quantum numbers
$n,l,s$ related by $n=l+s$. Since the problem involves only a magnetic
field, the solutions do not depend on the eigenvalues $\chi$. 

The solutions can be conveniently arranged in a matrix form 
\bea
   \Psi_p(x)=\sum_{\sigma=\pm 1}\psi_{p\sigma}(x)\Delta(\sigma)
   \label{solmat}
\eea
where
$
   \Delta(\sigma)\equiv{\mbox{diag}}\left\{\delta_{\sigma 1},
   \delta_{\sigma -1},\delta_{\sigma 1},\delta_{\sigma
   -1}\right\}
$
is a $4\times 4$ matrix and $x=(t,{\bf r})$.

The matrix in Eq.~(\ref{solmat}) is used to {\it rotate} the
two-point fermion Green's function between coordinate, $G(x,y)$ and
momentum spaces, ${\mathcal{G}}(k)$, as 
\bea
   G(x,y)=\sum_{n_k,s_k}\int d^2k_\parallel
   \Psi_k(x){\mathcal{G}}(k)\bar{\Psi}_k(y)\, ,
   \label{Gdiag}
\eea
where $k_\parallel=(k_0,0,0,k_3)$ and
$\bar{\Psi}_k=\gamma^0\Psi^\dagger_k\gamma^0$. The above expression
can be substituted into the equation relating the two-point fermion 
Green's function and the mass operator $M(x,y)$ in coordinate space,
namely 
\bea
   \gamma\cdot\Pi(x)G(x,y)\!-\!\!\int \!\!d^4x'M(x,x')G(x',y)
   =\delta^4(x-y)
   \label{Greeneq}
\eea
to find the explicit form for the function ${\mathcal{G}}$ in
momentum space, which is given by
\bea
   {\mathcal{G}}(k)=
   \frac{1}{\gamma\cdot k - \Sigma(k)}\, .
   \label{Gexplicitp}
\eea
In order to arrive at this equation, we have used the completeness of the
functions $\psi_{k\sigma}$ expressed in terms of $\Psi_k$ as
\bea
   \sum_{n_k,s_k}\int d^2k_\parallel
   \Psi_k(x)\bar{\Psi}_k(y)=\delta^4(x-y)
   \label{complete}
\eea
along with the properties
\bea
   \gamma\cdot\Pi(x)\Psi_k(x)&=&\Psi_k(x)(\gamma\cdot k)\nonumber\\
   \int d^4x'M(x,x')\Psi_k(x')&=&\Psi_k(x)\Sigma(k)\, ,
   \label{properties}
\eea
and the definition of the mass operator $\Sigma(k)$ in momentum space
\bea
   M(k,k')&\equiv&\int d^4xd^4x'\bar{\Psi}_k(x)M(x,x')\Psi_{k'}(x')
   \nonumber\\
   &=&\delta_{n_kn_{k'}}\delta_{s_ks_{k'}}
   \delta^2(k_\parallel-k'_\parallel)\Sigma(k)\, .
   \label{defMxp}
\eea

With the aid of Eqs.~(\ref{Gexplicitp})--(\ref{defMxp}) it is now
straightforward to transform the SDE for the mass
operator in the rainbow approximation from coordinate space, namely,
\bea
   M(x,x')=-ie^2\gamma^\mu G(x,x')\gamma^\nu D^{(0)}_{\mu\nu}(x-x')\, ,
   \label{SDcoord}
\eea 
to momentum space, which now reads
\begin{widetext}
\bea
   \delta_{n_pn_{p'}}\delta_{s_ps_{p'}}
   \delta^2(p_\parallel-p'_\parallel)\Sigma(p)=-ie^2\int d^4xd^4x'
   \sum_{n_k,s_k}\int d^2k_\parallel
   \bar{\Psi}_p(x)\gamma^\mu\Psi_k(x)\frac{D^{(0)}_{\mu\nu}(x-x')}
   {\gamma\cdot k-\Sigma(k)}
   \bar{\Psi}_k(x')\gamma^\nu\Psi_{p'}(x')\, ,
   \label{SDmom}
\eea
\end{widetext}
where the bare photon propagator is
\bea
   D^{(0)}_{\mu\nu}(x-x')
   =\int \frac{d^4q}{(2\pi)^4}\frac{e^{-iq\cdot
   (x-x')}}{q^2+i\epsilon}\Delta^{(0)}_{\mu\nu}(q)\, .
   \label{barephotx}
\eea

Having considered the dependence of the mass function $\Sigma(k)$ on
the structures in Eq.~(\ref{structures}), its remaining, most general
form can be written as
$
   \slsh{k}-\Sigma(k)={\mathcal{F}}^{-1}(k)
   \left[\slsh{k}-{\mathcal{M}}(k)\right]  
   \, , 
$
where, as in the case of vacuum, 
${\mathcal{F}}(k)$ and ${\mathcal{M}}(k)$ are called the wave function
renormalization and mass functions, respectively, in the presence of
the field. We work in the Landau gauge ($\xi=0$) where we
know that for vacuum $F=1$. Since we aim at a description for small magnetic
field strengths, we naturally expect ${\mathcal{F}}\sim 1$ in the Landau
gauge. Furthermore, let us work with
the ansatz that ${\mathcal{M}}(k)$ is proportional to the unit
matrix. Weakness of the magnetic field also implies that the bare vertex is 
a reasonable choice in the sense that Ward Identity is satisfied in
the Landau gauge up to a correction connected with the mass function
${\mathcal{M}}(k)$. This correction might be expected to be small
because ${\mathcal{M}}(k) \sim {\mathcal{M}}(p)$ for small values of momenta
as the mass function is practically a constant, and it falls off sharply
as $1/k$ for large momenta.

The self-consistent equation for the mass function is obtained by
considering the diagonal part $(n_p=n_{p'}\ s_p=s_{p'})$ and taking
the trace of Eq.~(\ref{SDmom}). 
The integrals over $x$ and $x'$ in Eq.~(\ref{SDmom}) are readily
performed. Having set $n_p=n_{p'}\ s_p=s_{p'}$, the integral over $x$
is just the complex conjugate of the one over $x'$. The first one is
found from the expression 
\begin{widetext}
\bea
   \int d^4x\psi^*_{p\sigma_p}(x)\psi_{k\sigma_k}(x)e^{-iq\cdot x}&=&
   \delta^2(p_\parallel-k_\parallel-q_\parallel)
   (-1)^{s_k-(\sigma_p+1)/2}
   \sqrt{\frac{(n_p-(\sigma_p+1)/2)!s_k!}{(n_k-(\sigma_k+1)/2)!s_p!}}
   \nonumber\\
   &&\times\left(\frac{q_\perp^2}{4\gamma}\right)^
   {(l_k-l_p)/2-(\sigma_k-\sigma_p)/4}e^{-q_{\perp}^2/4\gamma}
   L_{n_p-(\sigma_p+1)/2}^{n_k-n_p-(\sigma_k-\sigma_p)/2}(q_\perp^2/4\gamma)
   L_{s_k}^{s_p-s_k}(q_\perp^2/4\gamma)\, ,
   \label{firstint}
\eea
\end{widetext}
where $q_\perp = (0,q_1,q_2,0)$. Equation~(\ref{firstint})
is real and thus, integrating over $x$ and $x'$  in
Eq.~(\ref{SDmom}), results in the square of the right-hand side of
Eq.~(\ref{firstint}). The presence of the delta function in this last
equation, allows easy integration over $k_\parallel$. Gathering the
above described elements, we
obtain self-consistent equation for the mass function
\begin{widetext}
\bea
   {\mathcal{M}}(p_\parallel,n_p)&=&\frac{-ie^2}{2}\sum_{\sigma_k,\sigma_p=\pm
   1} 
   \sum_{n_k,s_k}\frac{s_p!s_k!}{(n_p-\frac{(\sigma_p+1)}{2})!
   (n_k-\frac{(\sigma_k+1)}{2})!}
   \int \frac{d^4q}{(2\pi)^4}\frac{e^{-q_\perp^2/2\gamma}}{q^2+i\epsilon}
   \frac{{\mathcal{M}}((p-q)_\parallel,n_k)}{(p-q)_\parallel^2-
   2eBn_k-{\mathcal{M}}^2((p-q)_\parallel,n_k)}
   \nonumber\\
   &&\hspace{-10mm}
   \left(\frac{q_\perp^2}{4\gamma}\right)^
   {l_k-l_p-\frac{(\sigma_k-\sigma_p)}{2}}\!\!
   \left[ 2 + \frac{1}{q^2}\left(q_\perp^2(1-\delta_{\sigma_p\sigma_k})
   -q_\parallel^2\delta_{\sigma_p\sigma_k}\right)\right]\!\!
   \left[L_{n_p-\frac{(\sigma_p+1)}{2}}^{n_k-n_p-
   \frac{(\sigma_k-\sigma_p)}{2}} 
   (q_\perp^2/4\gamma)\right]^2\!\!\!\!
   \left[L_{s_k}^{s_p-s_k}(q_\perp^2/4\gamma)\right]^2\, ,
\eea
\end{widetext}
where in the notation for the mass function we have emphasized the
breakdown of Lorentz invariance. We expect that
${\mathcal{M}}((p-q)_\parallel,n_k)$ should be 
independent of $s_k$ since the energy only depends on the
principal quantum number $n_k$. Furthermore we assume that
${\mathcal{M}}((p-q)_\parallel,n_k)$ is a slowly varying function of
$n_k$ and thus make the approximation
${\mathcal{M}}((p-q)_\parallel,n_k)\sim{\mathcal{M}}((p-q)_\parallel,n_k=0)$.
For consistency we consider the case $n_p=0$. Hereafter, we employ the
more convenient notation
${\mathcal{M}}(k_\parallel,n_k=0)\equiv{\mathcal{M}}(k_\parallel)$
for generic arguments of the mass function. With these
considerations the sum over $s_k$ can be computed by means 
of the result in Ref.~\cite{Gautam}. It is worth mentioning that after
summing over $s_k$, the resulting equation is the same as Eq.~(50) in
Ref.~\cite{Leung} when considering the case $n_k=0$, which corresponds to the
strong field limit. 

In the situation where the magnetic field is weak, we expand 
$
   [(p-q)_\parallel^2-2eBn_k-{\mathcal{M}}^2((p-q)_\parallel)]^{-1}
$
as a geometric series in powers of $eB$. The remaining sum over $n_k$
can be performed also by resorting to Ref.~\cite{Gautam} yielding,
after a Wick rotation
\begin{widetext}
\bea
   {\mathcal{M}}(p_\parallel)&\simeq&\frac{\alpha}{4\pi^3}\int d^4q
   \frac{{\mathcal{M}}((p-q)_\parallel)}{q^2[(p-q)_\parallel^2+q_\perp^2
   +{\mathcal{M}}^2((p-q)_\parallel)]}\left\{3 + 
   \left[\frac{4}{[(p-q)_\parallel^2+q_\perp^2
   +{\mathcal{M}}^2((p-q)_\parallel)]^2}\right.\right.\nonumber\\
   &-&\left.\left.\frac{6(6-q_\perp^2/q^2)q_\perp^2}
   {[(p-q)_\parallel^2+q_\perp^2
   +{\mathcal{M}}^2((p-q)_\parallel)]^3}+\frac{36q_\perp^4}
   {[(p-q)_\parallel^2+q_\perp^2
   +{\mathcal{M}}^2((p-q)_\parallel)]^4}\right](eB)^2
   \right\}\, , \label{SDEfinal}
\eea
\end{widetext}
keeping only the lowest order contribution in $eB$. Notice that, as
expected, the $s_p$ dependence of the mass function disappears on
carrying out the sum over $s_k$.

Solving the above equation
numerically is still not trivial, owing to the fact that
the unknown function ${\mathcal{M}}((p-q)_\parallel)$ within the
integral is Lorentz non-invariant. However, we can always expand it
out in powers of $(eB)^2$. Therefore, 
${\mathcal{M}}((p-q)_\parallel)={\mathcal{M}}_0(p-q)+ (eB)^2 {\mathcal{M}_1}$,
where ${\mathcal{M}_1}$ 
is responsible for breaking the Lorentz invariance of 
${\mathcal{M}}_0(p-q)$. 

\begin{figure}[t!] 
{\centering
{\epsfig{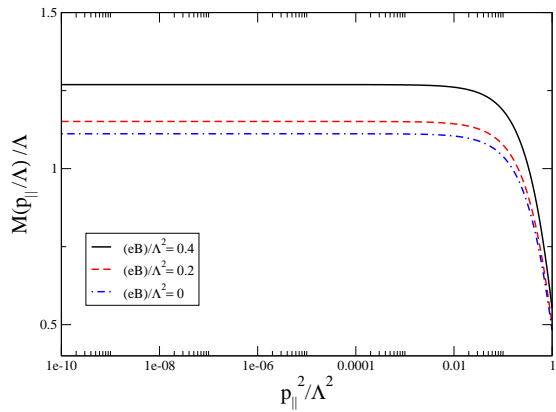}}
\par}
\caption{Mass function in the LLL for different values of the weak external
magnetic field for $\alpha=6.5\alpha_c$. }
\label{fig1}
\end{figure}

Consistently, we carry out the same expansion on the left hand side
of Eq.~(\ref{SDEfinal}).
As the Lorentz invariance should be restored for the leading terms,
we justifiably complete the momenta to achieve the same. This
filters out the vacuum result. To calculate the magnetic field
effect, we solve the integral equation for ${\mathcal{M}_1}$ 
obtained by comparing powers of $(eB)^2$. The results for
${\mathcal{M}}(p_\parallel /\Lambda)$ in the LLL are depicted in
Fig.~\ref{fig1}, scaled by the ultraviolet cut-off $\Lambda$. 
Note that the results have been shown for a value of 
the coupling significantly above the critical value only for
the magnetic field dependence to stand out. The same qualitative
behaviour persists even in the immediate vicinity of the critical
coupling,~\cite{footnote2}. 

\begin{figure}[t!] 
\vspace*{0.5cm}
{\centering
{\epsfig{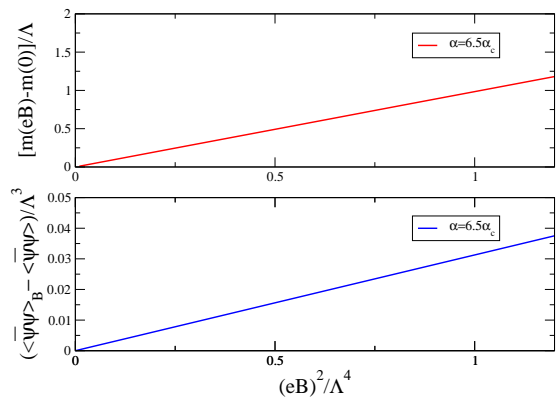}}
\par}
\caption{Magnetic contribution to the dynamically generated mass (upper graph)
  and condensate (lower graph) in the LLL as a function of $(eB)^2$ for
  $\alpha=6.5\alpha_c$.} 
\label{fig2}
\end{figure}

The dot-dashed
line corresponds to the vacuum. The effect of the external field is to
increase the dynamically generated mass, preserving the qualitative
features of the mass function profile. 
We have numerically verified that the critical value of the coupling 
is independent of the strength of the magnetic field, which is 
consistent with the findings of Ref.~\cite{Kikuchi}. To see the magnetic
contribution to the dynamically generated mass, we show in
Fig.~\ref{fig2} the difference $m(eB)-m(0)$, as a function of
$(eB)^2$, where $m$ is the dynamical 
fermion mass, namely, $m\equiv {\mathcal{M}}(0)$. Notice that this
difference grows linearly with $(eB)^2$. This is the same behaviour
as was observed for the NJL model in \cite{GusNJL}.
We also evaluate 
the condensate defined as $\langle\bar{\psi}\psi\rangle=i\ {\rm Tr}\ G(x,x)$. 
In the weak field limit, spacing between Landau levels
becomes small. To compute the condensate, the sum over these levels,
can be carried out by replacing $\sum_n$ with the integral $\int d^2
k_\perp/(2 \pi \ eB)$, along with the 
substitution $2eBn\rightarrow k_\perp^2$. The weak field contribution
to the condensate also turns out to be quadratic and with the above
mentioned substitutions, it owes itself entirely to the non Lorentz
invariant piece of the mass function in our computational set up. Its
behavior as a function of $(eB)^2$ is also shown in 
Fig.~\ref{fig2}.

In summary, we have shown that for $\alpha>\alpha_c$ the dynamically generated
mass increases quadratically with the magnetic field strength. As compared to
the strong field case, this is a four-fold dependence on the magnetic
field~\cite{Gusynin,Hong,Leung,Ferrer}. Such a dependence is similar to that
found in~\cite{GusNJL} for the NJL model. 
In the supercritical phase of QED, i.e, $\alpha >\alpha_{c}$, the gap equation of  Ref.~\cite{Kikuchi} can be solved  in the linearized approximation,  giving the same  quadratic dependence as we  have demonstrated through explicit numerical evaluation of the SDE in our setup~\cite{private}. 
It is ineteresting to note that
Farakos et. al.~\cite{Farakos} also report similar behaviour in
(2+1)-dimensional QED. Authors of this work employ the Schwinger proper time
method, neglecting the field dependent phase of the fermion
propagator. Therefore, a direct comparison with our findings is not
straightforward. 
Contrary to the widely studied case when the field is strong and the LLL
dominates, all the Landau levels 
should be taken into account in the weak field limit. This feature makes the
problem a difficult one and hence has been discussed  
less frequently in literature. Here we have shown that under plausible
assumptions about the behavior of the mass function, the sum over Landau
levels can be performed. The relaxation of some of the assumptions made is a
natural generalization of this work, along with the inclusion of a thermal
bath and the study of the gauge dependence of the results in the context of
the Ward identities~\cite{Ferrer2} and the Landau-Khalatnikov-Fradkin
transformations~\cite{BPR}. 

We acknowledge the valuable discussions with  V. de la Incera, E. Ferrer,
V. Gusynin, C.N. Leung, V.A. Miransky, C. Schubert and
A. S\'anchez. Support has been received in part by PAPIIT under Grant No.
IN107105 and CONACyT under Grant Nos. 40025-F and 46614-I.


\begin{thebibliography}{55}
\bibitem{vanish} Notice that the second and fourth structures in
  Eq.~(\ref{structures}) vanish identically for a constant magnetic field.
\end{thebibliography}
\end{document}